\title{An Explicit Construction of Quantum Expanders}

\author{Avraham Ben-Aroya \thanks{Schools of Computer Science, Raymond and Beverly
Sackler Faculty of Exact Sciences, Tel Aviv University, Tel Aviv
69978, Israel. Email: abrhambe@post.tau.ac.il. } \and Oded
Schwartz
\thanks{School of Computer Science,
Raymond and Beverly Sackler Faculty of Exact Sciences, Tel Aviv
University, Tel Aviv, Israel. Email: odedsc@tau.ac.il.} \and Amnon
Ta-Shma
\thanks{School of Computer Science,
Raymond and Beverly Sackler Faculty of Exact Sciences, Tel Aviv
University, Tel Aviv, Israel. Email: amnon@tau.ac.il. } }

\date{}

\documentclass [letterpaper,11pt]{article}
\usepackage{amsfonts}
\usepackage{amsmath,amsthm}

\newtheorem{lemma}{Lemma}[section]
\newtheorem{definition}[lemma]{Definition}
\newtheorem{prop}[lemma]{Proposition}

\newtheorem{claim}[lemma]{Claim}

\renewenvironment{proof}{{\bf Proof:\ }}{\hfill$\Box$\medskip}

\newcommand{\cH}{\mathcal{H}}

\newcommand{\ignore}[1]{}

\newcommand{\ol}[1]{\overline{#1}}
\newcommand{\lbar}{\ol{\lambda}}
\newcommand{\nI}{\tilde{I}}
\newcommand{\Tr}{\mathop{\rm Tr}\nolimits}
\newcommand{\norm}[1]{\left\|\,#1\,\right\|}
\newtheorem{theorem}{Theorem}
\newcommand{\zigzag}{\textcircled{z}}
\newcommand{\set}[1]{{\left\{#1\right\}}}
\newcommand{\tensor}{\otimes}
\newcommand {\ra} {\right \rangle}
\newcommand {\la} {\left  \langle}
\newcommand {\raabs} {\ra \right|}
\newcommand {\laabs} {\left| \la}
\newcommand{\ket}[1]{\left|#1\right\rangle}
\newcommand{\bra}[1]{\left\langle #1 \right|}
\newcommand{\braket}[2]{\left\langle #1\!\mid\! #2\right\rangle}
\newcommand{\ketbra}[2]{\ket{#1}\!\bra{#2}}
\newcommand{\Span}{\mathop{\rm Span}\nolimits}
\newcommand{\abs}[1]{\left|#1\right|}

\renewenvironment{proof}[1][]{\begin{trivlist}
\item[\hspace{\labelsep}{\bf\noindent Proof#1:\/}] }{\qed\end{trivlist}}
\renewcommand{\qed}{\hfill{\rule{2mm}{2mm}}}

\begin{document}
\maketitle

\begin{abstract}
Quantum expanders are a natural generalization of classical
expanders. These objects were introduced and studied by
\cite{bt:qed,hastings:qexpander,hastings:randomqexpander}. In this
note we show how to construct explicit, constant-degree quantum
expanders. The construction is essentially the classical Zig-Zag
expander construction of \cite{rvw:zigzag}, applied to quantum
expanders.
\end{abstract}


\section{Introduction}\label{intro}

Classical expanders are graphs of low degree and high connectivity.
One way to measure the expansion of a graph is through the second
eigenvalue of its adjacency matrix. This paper investigates the
quantum counterpart of these objects, defined as follows. For a
linear space ${\cal V}$ we denote by $L({\cal V})$ the space of
linear operators from ${\cal V}$ to itself.

\begin{definition}
\label{def:qunatum-degree} We say an admissible superoperator $G:
L({\cal V}) \to L({\cal V})$ is \emph{$D$-regular} if $G={1 \over
D} \sum_d G_d$, and for each $d \in [D]$, $G_d(X)=U_d X
U_d^\dagger$ for some unitary transformation $U_d$ over ${\cal
V}$.
\end{definition}

\begin{definition}
\label{def:qexpander} An admissible superoperator $G:L({\cal V})
\to L({\cal V})$ is a $(N, D,\lbar)$ quantum expander if
$\dim({\cal V})=N$, $G$ is $D$-regular and:

\begin{itemize}
\item
$G(\nI)=\nI$, where $\nI$ denotes the completely-mixed state.

\item
For any $\rho \in L({\cal V})$ that is orthogonal to $\nI$ (with
respect to the Hilbert-Schmidt inner product, i.e. $\Tr(\rho \nI) =
0$) it holds that $\norm{G(A)} \le \lbar \norm{A}$ (where
$\norm{X}=\sqrt{\Tr(X X^\dagger)}$).
%
\end{itemize}
A quantum expander is \emph{explicit} if $G$ can be implemented by
a quantum circuit of size polynomial in $\log(N)$.
\end{definition}

The notion of quantum expanders was introduced and studied by
\cite{bt:qed,hastings:qexpander,hastings:randomqexpander}. These
papers gave several constructions and applications of these
objects. The disadvantage of all the constructions given by these
papers is that each construction is either constant-degree or
explicit, but not both. In this paper we show how to construct
explicit quantum expanders of constant-degree. Our construction is
an easy generalization of the Zig-Zag expander construction given
in \cite{rvw:zigzag}.

\section{Preliminaries}

We denote by $\mathcal{H}_{N}$ the Hilbert space of dimension $N$.

For a linear space $\mathcal{V}$, we denote by $L(\mathcal{V})$ the
space of linear operators from $\mathcal{V}$ to itself. We use the
Hilbert-Schmidt inner product on this space, i.e. for $X, Y \in
L(\mathcal{V})$ their inner product is $\la X, Y \ra = \Tr (X
Y^\dagger)$. The inner product gives rise to a norm $\norm{X} =
\sqrt {\la X, X \ra} = \sqrt{\sum s_i(X)^2}$, where $\set{s_i(X)}$
are the singular values of $X$. Throughout the paper this is the
only norm we use.

We also denote by $U(\mathcal{V})$ the set of all unitary
operators on $\mathcal{V}$, and by $T(\mathcal{V})$ the space of
superopeartors on $\mathcal{V}$ (i.e. $T(\mathcal{V}) =
L(L(\mathcal{V}))$).

Finally, we denote by $\nI$ the identity operator normalized such
that $\Tr(\nI) = 1$. That is, $\nI$ denotes the completely mixed
state (on the appropriate space).

\section{Explicit constant-degree quantum expanders}

\subsection{The basic operations}
The construction uses as building blocks the following operations:
\begin{itemize}
    \item \textbf{Squaring:} For a superoperator $G \in T(\mathcal{V})$ we denote by $G^2$ the
    superoperator given by $G^2(X) = G(G(X))$ for any $X \in L({\cal V})$.

    \item \textbf{Tensoring:} For superoperators $G_1 \in T(\mathcal{V}_1)$ and
    $G_2 \in T(\mathcal{V}_2)$ we denote by $G_1 \tensor G_2$ the
    superoperator given by $(G_1 \tensor G_2)(X \tensor Y) = G_1(X) \tensor G_2(Y)$ for any $X \in L({\cal V}_1), Y \in L({\cal V}_2)$.

    \item \textbf{Zig-Zag product:} For superoperators $G_1 \in T(\mathcal{V}_1)$ and
    $G_2 \in T(\mathcal{V}_2)$ we denote by $G_1 \zigzag G_2$ their Zig-Zag product.
    A formal definition of this is given in Section \ref{sec:zigzag}. The only
    requirement is that $G_1$ is $\dim({\cal V}_2)$-regular.
\end{itemize}

\begin{prop}\label{prop:square}
If $G$ is a $(N, D, \lambda)$ quantum expander then $G^2$ is a $(N,
D^2, \lambda^2)$ quantum expander. If $G$ is explicit then so is
$G^2$.
\end{prop}

\begin{prop}\label{prop:tensor}
If $G_1$ is a $(N_1, D_1, \lambda_1)$ quantum expander and $G_2$ is
a $(N_2, D_2, \lambda_2)$ quantum expander then $G_1 \tensor G_2$ is
a $(N_1 \cdot N_2, D_1 \cdot D_2, \max(\lambda_1, \lambda_2))$
quantum expander. If $G_1$ and $G_2$ are explicit then so is $G_1
\tensor G_2$.
\end{prop}

\begin{theorem}\label{thm:zigzag}
If $G_1$ is a $(N_1, D_1, \lambda_1)$ quantum expander and $G_2$ is
a $(D_1, D_2, \lambda_2)$ quantum expander then $G_1 \zigzag G_2$ is
a $(N_1 \cdot D_1, D_2^2, \lambda_1 + \lambda_2 + \lambda_2^2)$
quantum expander. If $G_1$ and $G_2$ are explicit then so is $G_1
\zigzag G_2$.
\end{theorem}

The proofs of Propositions \ref{prop:square} and \ref{prop:tensor}
are trivial. The proof of Theorem \ref{thm:zigzag} is given in
Section \ref{sec:zigzag}.

\subsection{The construction}

The construction starts with some constant-degree quantum expander,
and iteratively increases its size via alternating operations of
squaring, tensoring and Zig-Zag products. The tensoring is used to
square the dimension of the superoperator. Then a squaring operation
improves the second eigenvalue. Finally, the Zig-Zag product reduces
the degree, without deteriorating the second eigenvalue too much.

Suppose $H$ is a $(D^8, D, \lambda)$ quantum expander. We define a
series of superoperators as follows. The first two superoperators
are $G_1 = H^2$ and $G_2 = H \tensor H$. For every $t > 2$ we define
$$ G_t = \left( G_{\lceil \frac{t-1}{2} \rceil} \tensor G_{\lfloor \frac{t-1}{2} \rfloor} \right)^2 \zigzag H .$$

\begin{theorem} \label{thm:qexapnder-construct}
For every $t>0$, $G_t$ is an explicit $(D^{8t},D^2,\lambda_t)$
quantum expander with $\lambda_t = \lambda + O(\lambda^2)$.
\end{theorem}

The proof of this Theorem for classical expanders was given in
\cite{rvw:zigzag}. The proof only relies on the properties of the
basic operations. Proposition \ref{prop:square}, Proposition
\ref{prop:tensor} and Theorem \ref{thm:zigzag} assure the required
properties of the basic operations are satisfied in the quantum case
as well. Hence, the proof of this theorem is identical to the one in
\cite{rvw:zigzag} (Theorem 3.3) and we omit it.

\subsection{The base superoperator} Theorem \ref{thm:qexapnder-construct}
relies on the existence of a good base superoperator $H$. In the
classical setting, the probabilistic method assures us that a good
base graph exists, and so we can use an exhaustive search to find
one. The quantum setting exhibits a similar phenomena:
\begin{theorem} \label{thm:random-qexpander}
(\cite{hastings:randomqexpander}) There exists a $D_0$ such that
for every $D > D_0$ there exist a $(D^8,D,\lambda)$ quantum
expander for $\lambda = \frac{4 \sqrt{D-1}}{D}$
\footnote{\cite{hastings:randomqexpander} actually shows that for
any $D$ there exist a $(D^8,D,(1+O(D^{-16/15} \log D)) \frac{2
\sqrt{D-1}}{D})$ quantum expander.}.
\end{theorem}
We will use an exhaustive search to find such a quantum expander.
To do this we first need to transform the searched domain from a
continuous space to a discrete one. We do this by using a net of
unitary matrices, $S \subset U(\mathcal{H}_{D^8})$. $S$ has the
property that for any unitary matrix $U \in U(\mathcal{H}_{D^8})$
there exists some $V_U \in S$ such that
$$ \sup_{\norm{X} = 1} \norm{UXU^\dagger - V_{U}X V_{U}^\dagger} \le \lambda .$$
It is not hard to verify that indeed such $S$ exists, with size
depending only on $D$ and $\lambda$. Moreover, we can find such a
set in time depending only on $D$ and $\lambda$ \footnote{One way
to see this is using the Solovay-Kitaev theorem (see, e.g.,
\cite{dn:skalgorithm}). The theorem assures us that, for example,
the set of all the quantum circuits of length $O(\log^4
\epsilon^{-1})$ generated only by Hadamard and Tofolli gates give
an $\epsilon$-net of unitaries. The accuracy of the net is
measured differently in the Solovay-Kitaev theorem, but it can be
verified that the accuracy measure we use here is roughly
equivalent.}. 

Suppose $G$ is a $(D^8,D,\lambda)$ quantum expander, $G(X) =
\frac{1}{D} \sum_{i=1}^D U_i X U_i^\dagger$. We denote by $G'$ the
superoperator $G'(X) = \frac{1}{D} \sum_{i=1}^D V_{U_i} X
V_{U_i}^\dagger$. Let $X \in L(\mathcal{H}_{D^8})$ be orthogonal to
$\nI$. Then:
\begin{align*}
 \norm{G'(X)} &= \norm{\frac{1}{D} \sum_{i=1}^D V_{U_i} X
V_{U_i}^\dagger} \le \norm{\frac{1}{D} \sum_{i=1}^D U_i X
U_i^\dagger} + \frac{1}{D} \sum_{i=1}^D \norm{U_i X U_i^\dagger -
V_{U_i} X V_{U_i}^\dagger} \\
&\le \norm{G(X)} + \lambda \norm{X} \le 2\lambda \norm{X}.
\end{align*}
Hence, $G'$ is a $(D^8,D,\frac{8 \sqrt{D-1}}{D})$ quantum expander
\footnote{We can actually get an eigenvalue bound of $(1+\epsilon)
\frac{2 \sqrt{D-1}}{D}$ for an arbitrary small $\epsilon$ on the
expense of increasing $D_0$.}. This implies that we can find a
good base superoperator in time which depends only on $D$ and
$\lambda$.

\section{The Zig-Zag product}
\label{sec:zigzag}

Suppose $G_1,G_2$ are two superoperators, $G_i \in T(\cH_{N_i})$,
and $G_i$ is a $(N_i,D_i,\lambda_i)$ quantum expander. We further
assume that $N_2=D_1$. $G_1$ is $D_1$--regular and so it can be
expressed as $G_1(X)={1 \over D_1} \sum_d U_d X U_d^\dag$ for some
unitaries $U_d \in U(\cH_{N_1})$. We lift the ensemble $\set{U_d}$
to a superoperator $\dot{U} \in L(\cH_{N_1} \tensor \cH_{D_1})$
defined by:

\begin{eqnarray*}
\dot{U}(\ket{a} \tensor \ket{b}) &=& U_b \ket{a} \tensor \ket{b},
\end{eqnarray*}
and we define $\dot{G_1} \in T(\cH_{N_1} \tensor \cH_{D_1})$ by
$\dot{G_1}(X)=\dot{U}X \dot{U}^\dag$.

\begin{definition}
Let $G_1, G_2$ be as above. The Zig-Zag product, $G_1 \zigzag G_2
\in T(\mathcal{H}_{N_1} \tensor \mathcal{H}_{D_1})$ is defined to be
$(G_1 \zigzag G_2)X = (I \tensor G_2) \dot{G_1} (I \tensor
G_2^\dagger) X$.
\end{definition}

We claim:

\begin{prop}
\label{prop:zigzag} For any $X,Y \in L(\cH_{N_1} \tensor
\cH_{D_1})$ such that $X$ is orthogonal to the identity operator
we have:

\begin{eqnarray*}
~|~ \la ~ G_1 \zigzag G_2 X ~,~ Y \ra ~| & \le &
f(\lambda_1,\lambda_2) \norm{X} \cdot \norm{Y}
\end{eqnarray*}
where $f(\lambda_1,\lambda_2)=\lambda_1 + \lambda_2 +
\lambda_2^2$.
\end{prop}

And as a direct corollary we get:

\newtheorem*{thma}{Theorem~\ref{thm:zigzag}}
\begin{thma}
If $G_1$ is a $(N_1, D_1, \lambda_1)$ quantum expander and $G_2$ is
a $(D_1, D_2, \lambda_2)$ quantum expander then $G_1 \zigzag G_2$ is
a $(N_1 \cdot D_1, D_2^2, \lambda_1 + \lambda_2 + \lambda_2^2)$
quantum expander. If $G_1$ and $G_2$ are explicit then so is $G_1
\zigzag G_2$.
\end{thma}


\begin{proof}
Let $X$ be orthogonal to $\nI$ and let $Y = (G_1 \zigzag G_2)X$.
By Proposition \ref{prop:zigzag} $\norm{Y}^2 \le
f(\lambda_1,\lambda_2) \norm{X} \cdot \norm{Y}$. Equivalently,
$\norm{(G_1 \zigzag G_2)X} \le f(\lambda_1,\lambda_2) \norm{X}$ as
required.

The explicitness of $G_1 \zigzag G_2$ is immediate from the
definition of the Zig-Zag product.
\end{proof}

We now turn to the proof of Proposition \ref{prop:zigzag}. We adapt
the proof given in \cite{rvw:zigzag} for the classical case to the
quantum setting. For that we need to work with linear operators
instead of working with vectors. Consequently, we replace the vector
inner-product used in the classical proof with the Hilbert-Schmidt
inner product on linear operators, and replace the Euclidean norm on
vectors, with the $\Tr(X X^\dag)$ norm on linear operators.
Interestingly, the same proof carries over to this generalized
setting. One can get the proof below by simply going over the proof
in \cite{rvw:zigzag} and doing the above translation. We provide the
details here for completeness.

\begin{proof}[ of Proposition \ref{prop:zigzag}]
We first decompose the space $L(\cH_{N_1} \tensor \cH_{D_1})$ to

\begin{eqnarray*}
W^{||} &=& \Span \set{\sigma \tensor \widetilde{I} ~|~ \sigma \in L(\cH_{N_1})} \mbox{ and,}\\
W^{\perp} &=& \Span \set{\sigma \tensor \tau ~|~ \sigma \in
L(\cH_{N_1})~,~\tau \in L(\cH_{D_1})~,~ \la \tau , \widetilde{I} \ra
= 0 }.
\end{eqnarray*}

Decompose $X$ to $X = X^{||}+X^{\perp}$, where $X^{||} \in W^{||}$
and $X^{\perp}\in W^{\perp}$, and similarly $Y=Y^{||}+Y^{\perp}$.
By definition,
$$|\la G_1 \zigzag G_2 X, Y \ra| =  |\la (I \tensor G_2) \dot{G_1} (I \tensor G_2^\dagger)X , Y\ra|
= |\la \dot{G_1} (I \tensor G_2) (X^{||}+X^{\perp}) , (I \tensor
G_2)(Y^{||}+Y^{\perp})\ra|.$$ Opening to the four terms and pushing
the absolute value inside, we see that

\begin{eqnarray*}
|\la G_1 \zigzag G_2 X, Y \ra| & \le & |\la \dot{G_1} (I \tensor
G_2) X^{||} , (I \tensor G_2)Y^{||} \ra | + |\la \dot{G_1} (I
\tensor G_2) X^{||} , (I \tensor
G_2)Y^{\perp}  \ra |  +\\
&  & |\la \dot{G_1} (I \tensor G_2) X^{\perp} , (I \tensor
G_2)Y^{||} \ra| + | \la \dot{G_1} (I \tensor G_2) X^{\perp},(I
\tensor G_2)Y^{\perp} \ra| \\
& = & |\la \dot{G_1} X^{||} , Y^{||}  \ra| +| \la \dot{G_1} X^{||} ,
(I \tensor G_2)Y^{\perp} \ra| + \\
&  & |\la \dot{G_1} (I \tensor G_2) X^{\perp} , Y^{||} \ra | +| \la
\dot{G_1} (I \tensor G_2) X^{\perp} , (I \tensor G_2)Y^{\perp} \ra |
\end{eqnarray*}

Where the last equality is due to the fact that $I \tensor G_2$ is
identity over $W^{||}$ (since $G_2(\nI)=\nI$). In the last three
terms we have $I \tensor G_2$ acting on an operator from
$W^\perp$. As expected, when this happen the quantum expander
$G_2$ shrinks the operator. Formally,

\begin{claim}\label{cl:zigzag:perp}
For any $Z \in W^{\perp}$ we have $\norm{ (I \tensor G_2)Z} \le
\lambda_2 \norm{Z}$.
\end{claim}

We defer the proof for later. Having the claim we see that, e.g., $|
\la \dot{G_1} X^{||} , (I \tensor G_2)Y^{\perp} \ra| \le \norm{
\dot{G_1} X^{||} } \cdot \norm{(I \tensor G_2) Y^{\perp}} \le
\lambda_2 \norm{X^{||}} \cdot \norm{Y^{\perp}}$. Similarly, $|\la
\dot{G_1} (I \tensor G_2) X^{\perp} , Y^{||} \ra | \le \lambda_2
\norm{X^{\perp}} \cdot \norm{Y^{||}}$ and $| \la \dot{G_1} (I
\tensor G_2) X^{\perp} , (I \tensor G_2)Y^{\perp} \ra | \le
\lambda_2^2 \norm{X^{\perp}} \norm{Y^{\perp}}$.

To bound the first term, we notice that on inputs from $W^{||}$ the
operator $\dot{G_1}$ mimics the operation of $G_1$ with a random
seed. Formally,

\begin{claim}\label{cl:zigzag:parallel}
For any $A \in W^{||}$ orthogonal to the identity operator and any
$B \in W^{||}$ we have $| \la \dot{G_1} A, B \ra | \le \lambda_1
\norm{A} \cdot \norm{B}$.
\end{claim}

We again defer the proof for later. Having the claim we see that
$|\la \dot{G_1} X^{||} , Y^{||} \ra | \le \lambda_1 \norm{X^{||}}
\cdot \norm{Y^{||}}$. Denoting  $p_i = {{\norm{\rho_i^{||}}} \over
{\norm{\rho_i}}}$ and $q_i = {{\norm{\rho_i^{\perp}}} \over
{\norm{\rho_i} }}$ (for $i=1,2$, $\rho_1=X$ and $\rho_2=Y$) we see
that $p_i^2+q_i^2=1$, and,

\begin{eqnarray*}
| \la (G_1 \zigzag G_2)X, Y \ra | & \le & ( p_1 p_2 \lambda_1 + p_1
q_2\lambda_2 + p_2 q_1\lambda_2 + q_1 q_2 \lambda_2^2 ) \norm{X}
\cdot \norm{Y}
\end{eqnarray*}

Elementary calculus  now shows that this is bounded by
$f(\lambda_1,\lambda_2) \norm{X} \cdot \norm{Y}$.
\end{proof}

We still have to prove the two claims:

\begin{proof}[ of Claim \ref{cl:zigzag:perp}]
$Z$ can be written as $Z=\sum_{i} \sigma_i \tensor \tau_i$, where
each $\tau_i$ is perpendicular to $\nI$ and $\set{\sigma_i}$ is an
orthogonal set. Hence,
$$\norm{ (I \tensor G_2)Z} = \norm{\sum_{i} \sigma_i \tensor G_2(\tau_i)} \le
\sum_{i} \norm{\sigma_i \tensor G_2(\tau_i)} \le \sum_{i}
\lambda_2 \norm{\sigma_i \tensor \tau_i} = \lambda_2 \norm{Z}.$$
\end{proof}

And,

\begin{proof}[ of Claim \ref{cl:zigzag:parallel}]
Since $A,B \in W^{||}$, they can be written as
$$ A = \sigma \tensor \nI = \frac{1}{D_1} \sum_i \sigma \tensor \ketbra{i}{i}$$
$$ B = \eta   \tensor \nI = \frac{1}{D_1} \sum_i \eta   \tensor \ketbra{i}{i}.$$
Moreover, since $A$ is perpendicular to the identity operator, it
follows that $\sigma$ is perpendicular to the identity operator on
the space $L(\mathcal{H}_{N_1})$. This means that applying $G_1$ on
$\sigma$ will shrink it by at least a factor of $\lambda_1$.

Considering the inner product
\begin{align*}
|~\la \dot{G_1}A,B \ra | &= \frac{1}{D_1^2} \abs{\sum_{i,j} \Tr
\left( \left( (U_i \sigma U_i^\dagger)\tensor \ketbra{i}{i}
\right) \left( \eta \tensor
\ketbra{j}{j} \right)^\dagger \right)} \\
&= \frac{1}{D_1^2} \abs{\sum_{i,j} \Tr \left( (U_i \sigma
U_i^\dagger \eta^\dagger)\tensor \ket{i}\braket{i}{j}\bra{j} \right)} \\
&= \frac{1}{D_1^2} \abs{\sum_{i} \Tr \left( (U_i \sigma
U_i^\dagger \eta^\dagger)\tensor \ketbra{i}{i} \right)} \\
&= \frac{1}{D_1^2} \abs{\sum_{i} \Tr \left( U_i \sigma
U_i^\dagger \eta^\dagger \right)}  \\
&= \frac{1}{D_1} \abs{\Tr \left( \left( \frac{1}{D_1} \sum_{i} U_i
\sigma U_i^\dagger \right) \eta^\dagger \right)}\\
&= \frac{1}{D_1} \laabs G_1(\sigma),\eta \raabs \le
\frac{\lambda_1}{D_1}\norm{\sigma} \cdot \norm{\eta} = \lambda_1
\norm{A} \cdot \norm{B} ,
\end{align*}
where the inequality follows from the expansion property of $G_1$
(and Cauchy-Schwartz).
\end{proof}

\bibliographystyle{plain}
\bibliography{qc}

\end{document}